\documentclass[aps,prl,twocolumn,floats,floatfix,showpacs,ansmath,amssymb,superscriptaddress]{revtex4}
\usepackage{graphicx}
\usepackage{bm}

\begin{document}
%
%
\newcommand{\be}{\begin{equation}}
\newcommand{\ee}{\end{equation}}
\newcommand{\bea}{\begin{eqnarray}}
\newcommand{\eea}{\end{eqnarray}}
\newcommand{\beann}{\begin{eqnarray*}}
\newcommand{\eeann}{\end{eqnarray*}}
\newcommand{\bma}{\begin{array}{cc}}
\newcommand{\ema}{\end{array}}
\newcommand{\fr}{\frac}
\newcommand{\ra}{\rangle}
\newcommand{\la}{\langle}
\newcommand{\li}{\left}
\newcommand{\re}{\right}
\newcommand{\ri}{\right}

\newcommand{\uarr}{\uparrow}
\newcommand{\darr}{\downarrow}
\newcommand{\df}{\stackrel{\rm def}{=}}
\newcommand{\nn}{\nonumber}
\newcommand{\dpl}{\displaystyle}

\newcommand{\alp}{\alpha}
\newcommand{\sig}{\sigma}
\newcommand{\eps}{\epsilon}
\newcommand{\xsi}{\xi}
\newcommand{\lam}{\lambda}
\newcommand{\ny}{\nu}

%
%
\title{Classical Phase Space Revealed by Coherent Light}
\author{Tomoko Tanaka}
%
%
%
\affiliation{
Department of Nonlinear Science, ATR Wave Engineering Laboratories, 2-2-2 Hikaridai, Seika-cho, Soraku-gun, Kyoto 619-0228, Japan}
\author{Martina Hentschel}
\affiliation{
Department of Nonlinear Science, ATR Wave Engineering Laboratories, 2-2-2 Hikaridai, Seika-cho, Soraku-gun, Kyoto 619-0228, Japan}
\affiliation{
Institut f\"ur Theoretische Physik, Universit\"at Regensburg, D-93040 Regensburg, Germany}
\thanks{Present address: MPIPKS Dresden, N\"othnitzer Str. 38, D-01187 Dresden, Germany}
\author{Takehiro Fukushima}
\affiliation{
Department of Nonlinear Science, ATR Wave Engineering Laboratories, 2-2-2 Hikaridai, Seika-cho, Soraku-gun, Kyoto 619-0228, Japan}
\affiliation{
Department of Communication Engineering, Okayama Prefectural University, Soja 719-1197, Japan}

\author{Takahisa Harayama}
\affiliation{
Department of Nonlinear Science, ATR Wave Engineering Laboratories, 2-2-2 Hikaridai, Seika-cho, Soraku-gun, Kyoto 619-0228, Japan}

\date{\today}

%
%
\begin{abstract}
{We study the far field characteristics of oval-resonator laser
diodes made of an AlGaAs/GaAs quantum well. The resonator shapes are
various oval geometries, thereby probing chaotic and mixed
classical dynamics. The far field pattern shows a pronounced fine
structure that strongly depends on the cavity shape. Comparing the
experimental data with ray-model simulations for a Fresnel
billiard yields 
convincing agreement for all geometries and reveals
the importance of the underlying classical phase space for the
lasing characteristics.}
\end{abstract}
\maketitle


{\it Introduction} --- In the past two decades, quantum chaos has
proven to be a successful concept in understanding,
characterizing, and predicting the behavior of mesoscopic systems
\cite{stoeckmann,sohnetal}. Originally used to study the quantum
analogue of classically chaotic hard-wall billiard model systems,
it also explains the behavior of realistic systems of various
shapes and character: The statistics of Coulomb-blockade peaks in
quantum dots \cite{sohnetal}, the geometry dependence of the weak
localization peak \cite{weakloc}, or the level statistics of
microwave billiards \cite{microwavebill}. The studies on quantum
chaos have mainly focused on the quest for universality from the
view-point of statistical physics. Therefore, despite the evident
importance of the system's underlying classical phase space for
the behavior of the quantum or wave mechanical analogue (based on
the analogy between Schr\"odinger and Helmholtz equation
\cite{stoeckmann}), its specific structure cannot be reconstructed
from the traces it leaves in typical observables like energy level
or wave function statistics.

In this Letter we shall see that detailed information about the
classical phase space can, however, be extracted from the far
field radiation characteristics of oval-microcavity laser diodes.
The observed far
field pattern (FFP) depends very sensitively on the shape, that is,
on the system's underlying classical phase space, and we
convincingly support this idea by numerical simulations.
It is interesting to note that information about the classical phase space is
revealed by the {\it coherent} 
light emanating from the {\it lasing} microcavity.

\begin{figure}[tb]
\includegraphics[width=5.5cm, angle = -90]{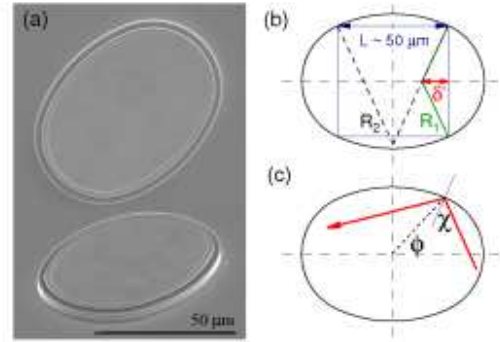}
\vspace*{-2mm}
\caption{(Color online)
An oval microlaser with $\delta\equiv 2 \delta^{\prime} / L
$=0.45. (a) Scanning Electron Micrograph, (b) geometry, and (c)
phase-space coordinates.
  }
\label{fig1}
\end{figure}

\begin{figure}[t]
\hspace*{-3mm}
\includegraphics[width=7.5cm]{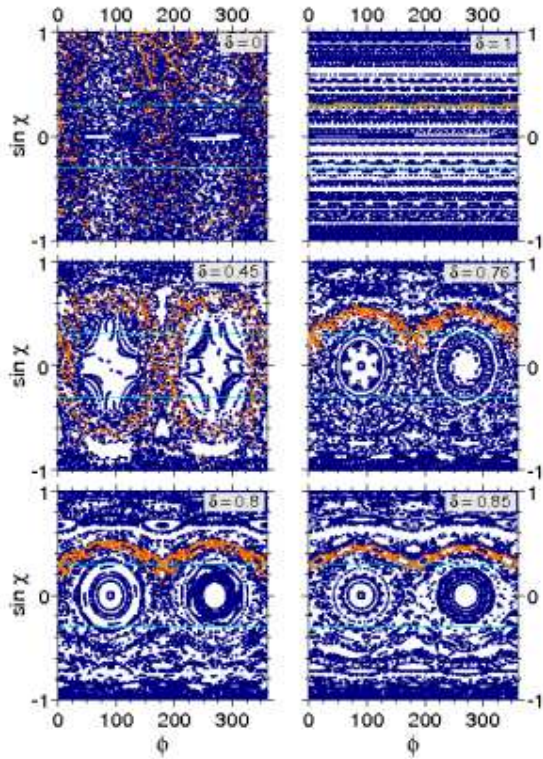}
\caption{(Color online) Poincar\'{e} surface of section (SOS)
illustrating the classical
phase space of hard-wall oval billiards. The critical lines $|\sin
\chi| =  1/n$ are shown by light-blue lines.
In the top raw, the limiting cases of a chaotic ($\delta=0$,
stadium) and an integrable ($\delta=1$, disk) phase space are
shown; for $\delta$=0.45, 0.76, 0.8, and 0.85 the phase space is
mixed. To further visualize the differences in the phase-space
structure, the evolution of 50 rays with random initial conditions
in the interval $0\leq \phi \leq 0.1$ and $0.28 \leq \sin \chi
\leq 0.32$ is followed for 25 reflections (superposed lighter
points).
  }
\label{newfig2}
\end{figure}

The oval-billiard family is a well-known model system with
interesting properties \cite{Benettin,Wisdom,Makino} in which each
oval shape is characterized by a shape parameter $\delta \equiv 2
\delta^{\prime} / L$, cf.~Fig.~\ref{fig1}(b). The curvature is, by construction [see
Fig.~\ref{fig1}(b) and, e.g., Ref.~\onlinecite{Makino} for
instructions], discontinuous. This was found to be crucial for
understanding 
the phase-space evolution upon
variation of $\delta$ \cite{Wisdom}. Increasing $\delta$, the
phase space changes from fully chaotic ($\delta=0$, stadium
shape), via mixed ($0 < \delta < 1$, various oval shapes) to
integrable ($\delta=1$, disk). Figure \ref{newfig2} shows the
phase space of the chaotic and integrable limit and the mixed
phase spaces for $\delta \in$ [$0.45$,$0.85$] that consist of
chaotic regions as well as regular islands \cite{Fukushima}. The
simultaneous experimental and theoretical investigation of a
$\delta$-dependent mixed phase space has not, to the best of our
knowledge, been done before and is at the core of the present
study.

To understand the dynamics of the system, let us consider a light
ray hitting the resonator boundary at polar angle $\phi$ under an
angle of incidence $\chi$, cf.~Fig.\ref{fig1}(c), such that total
internal reflection takes place ($|\sin
\chi| \geq 1/n$). In the circular cavity ($\delta=1$), the light
ray will remain confined inside the cavity forever by means of
conservation of angular momentum, and form whispering gallery mode
(WGM) tori. To a certain extent, this scenario will remain valid
even if $\delta$ is slightly decreased. The invariant WGM tori are
then perturbed to Kolmogorov-Arnold-Moser tori, and the phase
space becomes mixed. Note that for $\delta \geq $0.76 the chaotic
regions are disconnected (the chaotic sea is split into two
disjoined regions) whereas for $\delta<$0.76 a
singly connected chaotic layer penetrates the phase space,
bypassing the embedded regular islands \cite{Wisdom,Makino},
cf.~Fig.~\ref{newfig2}. The smaller $\delta$ the more chaotic the
cavity becomes, reaching ergodicity in the limit $\delta=0$.

The above mentioned properties of ray-dynamics in oval billiards
imply in particular that when $\delta$ exceeds values of $\sim
0.76$, light rays are trapped in the disjoined chaotic regions
that lie above the critical line $|\sin \chi| = 1/n$, unable to
leave the cavity. Accordingly, the emitted light intensity is
expected to decrease around $\delta \sim 0.76$ because the lasing 
modes will be strongly confined inside the cavity and leakage 
will be evanescent.

In order to elucidate if this property of classical phase
space could be detected in the emitted light,
we actually fabricated these oval-shaped semiconductor laser
diodes for various values of $\delta$, one of which is shown in Fig.~1(a).

\begin{figure}[t]
\hspace*{-5mm}
\includegraphics[width=6.5cm]{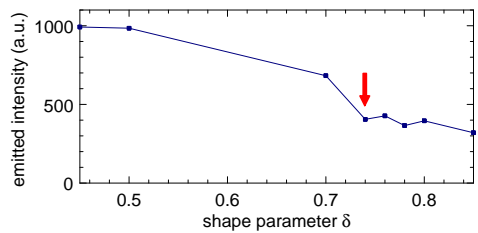}
\caption{
Total emitted intensity of lasing microcavities vs.~shape 
parameter $\delta$. Note the intensity drop (marked by arrow)
around $\delta \sim 0.76$ caused by the appearance of a transversal
barrier in phase. 
  }
\label{newfig3}
\end{figure}

{\it System} --- The resonators were made of metal organic vapor
phase epitaxy grown graded index seperate confinement AlGaAs/GaAs
single-quantum well wafer with $n$=3.3 the effective index of
refraction. The typical system length is $L \sim$ 50 $\mu m$. Applying the
reactive-ion-beam etching technique allows one to realize specific
oval cavities characterized by a 
shape parameter $\delta$ with extremely smooth and vertical
boundaries, see Fig.~1(a); the surface roughness is less than 1/10
of the lasing wavelength $\lambda \sim 850$ nm. Lasing operation
of these devices was achieved at room temperature by pumping with
a pulsed injection current of 500 ns width at a 1 kHz repetition
rate.

{\it Experiments} --- In Fig.~\ref{newfig3} we plot
the total (polar-angle integrated) far field output power,
measured at an injection current of four times the threshold current,
as function of the shape parameter $\delta$.
The pronounced drop in the emitted light intensity around $\delta
\lesssim 0.76$ (marked by arrow) is a direct consequence of the
above-mentioned appearance of a transversal barrier in classical
phase space and the change of the lasing modes from 
refractive modes into evanescent, WGM-type modes.
This dip is also a clear sign that not individual modes, but the
underlying classical phase space as a whole (representing a wealth
of modes \cite{remark_FSR}) determines the behaviour of the system.

\begin{figure}[tb]
\hspace*{-3mm}
\includegraphics[width=7.5cm]{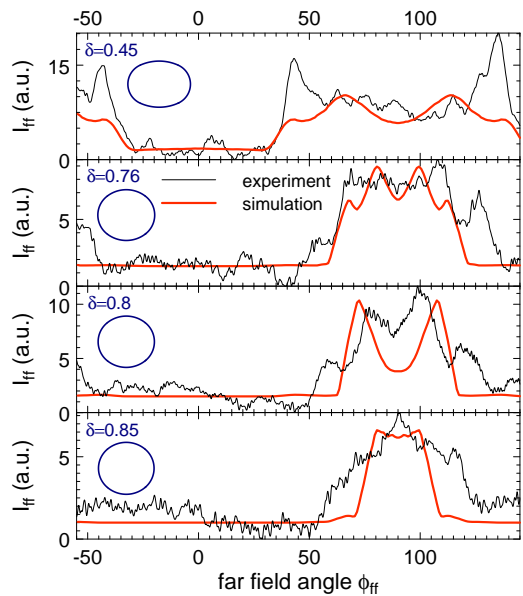}
\caption{(Color online) Experimentally observed FFPs
(far field intensity $I_{\rm ff}$ vs.~angle
$\phi_{\rm ff}$, black curves; same injection currents as in Fig.~\ref{newfig3} were used)
for various
shape parameters $\delta$ (see insets). The ray-model
simulations are superimposed (red curves); all data
was averaged over a 9$^\circ$ window. 
The simulation data is offset by the
average residual background of the experimental data,
normalization is with respect to
the peak heights.
Apart from the side peaks, the agreement
experiment-theory is convincing and the drastic changes in the FFP
upon tiny changes in the shapes are nicely reflected in the simulation.
  }
\label{newfig4}
\end{figure}

In the following, we will shift our focus from the total output
power to the angular distribution of the emitted radiation in the
far field, i.e., the FFPs. The experimental FFPs of
oval lasing diodes with various $\delta$ are
shown in Fig.~\ref{newfig4} (black curves). FFPs were measured in
a distance of $d$ = 70 mm by scanning a photo-detector with a
window of $w$ = 11 mm width around the laser diode covering an
angular range of 200$^\circ$ in $\phi_{\rm ff}$ with a resolution
$\Delta \phi_{\rm ff} = 2 \arctan( w / 2 d ) \sim 9^\circ$.

The red curves in Fig.~\ref{newfig4} are the results of
numerical, ray-model based simulations (explained in more detail
below) that convincingly support the experimental results.
A uniform background was
substracted in all cases in order to exclude effects of spontaneous emission
that is assumed to be isotropic; the (absolute) minimum of the
experimental curves is set to zero by definition.

\begin{figure}[t]
\hspace*{-2mm}
\includegraphics[width=8.5cm]{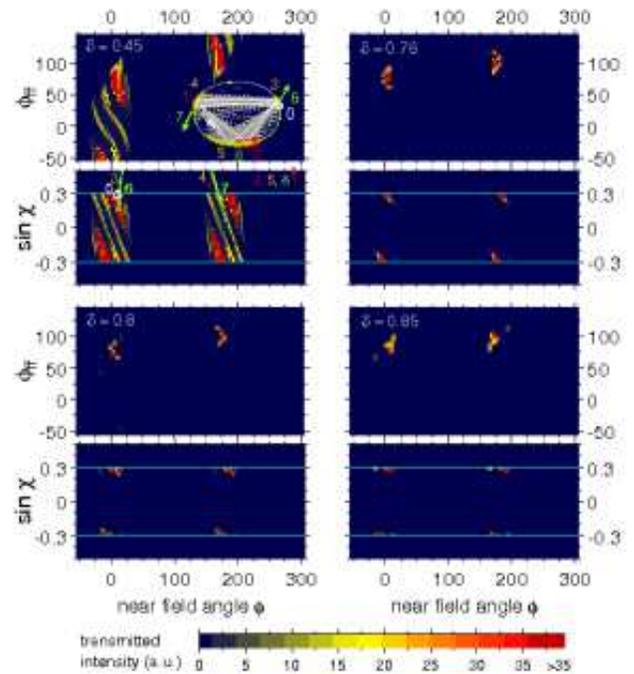}
\caption{(Color online) From classical phase space to the FFPs of 
microcavities (ray-model based simulation). The transmitted
Fresnel intensity (color scale) is shown in the near field (lower
panels, sine of angle of incidence $\chi$ vs.~polar angle angle
$\phi$ at transmission) and far field (upper panels, resulting
far field angle $\phi_{\rm ff}$
vs.~polar angle $\phi$ of ray
origin). The region of violated total internal reflection is
enclosed by light lines in the lower panels. The intensity distribution in the
lower panels corresponds well to the hard-wall results (cf.~the lighter points
in Fig.~\ref{newfig2}); their translation into
the FFP in the upper panels and in Fig.~\ref{newfig4} is evident.
The inset for $\delta=0.45$ illustrates how the details of the
observed structure are related to shape-specific orbits as the one
shown (see text).
  }
\label{newfig5}
\end{figure}

{\it Model} ---
The theoretical model used here is based on the ray picture. This
is motivated by the large size parameter $\sim 2 \pi n L/\lambda
\sim 1000$ of the experimental system; the ray model can be expected
to yield reliable results in this regime.
\cite{Fukushima,raymodel_ok} A large number (250000) of rays is
started with unit amplitude and random initial conditions covering
the whole phase space. The dynamics of each ray is governed by
Snell's and Fresnel's law \cite{remark_pol} that complement the
condition of specular reflection. The far field radiation
characteristics is reconstructed from rays in the steady
phase-space distribution \cite{Fukushima,steady_phsp_distr}, an
invariant object uniquely related to the underlying phase-space
structure and characterized by an exponential decay of the
internal light intensity vs.~trajectory length. The initial
transient regime where the internal intensity decays faster than
exponentially due to the presence of, e.g., bouncing ball orbits,
is discarded \cite{Fukushima}. That the linear ray-model works so
well for describing non-linear, lasing light is related to the
large size-parameter and the fact that not a single, but a
multitude of modes is lasing. \cite{remark_FSR}

{\it Results} --- The observed FFPs exhibit a clear and prominent
minimum-maximum structure that is remarkably well reproduced by
the ray-model simulations for {\it all} $\delta$,
cf.~Fig.~\ref{newfig4}. Apart from the edges of the broad maxima
where side peaks are sometimes missing,
the simulated FFPs capture the decreasing
width and fine structure of the central part, in
particular its extreme shape-sensitivity, in a semi-quantitative
manner.

For all $\delta$ considered here, the preferred emission direction
coincides with the shorter cavity axis ($\phi_{\rm ff} \sim 90^\circ$). 
Note that this direction is shifted by 90$^\circ$ in comparison
with the stadium shaped microcavity laser \cite{Fukushima} due to
dynamical eclipsing \cite{Noeckel}: For
$\delta>0$, regular (bouncing ball) islands around the
stable fixed points $\sin \chi=0$ and $\phi=0^\circ, 180^\circ$ 
impose constraints on all other orbits.  Light rays will still
emerge mainly from the regions of highest curvature, but
nearly {\it tangentially}, i.e., in direction of the
shorter cavity axis, cf.~also Figs.~\ref{newfig2} and
\ref{newfig5}.

In the remaining part of this Letter, we will focus
on the fine structure of the FFP maxima and show how the
underlying classical phase space expresses itself in this
signature, cf.~Figs.~\ref{newfig4} and \ref{newfig5}.
That the ray-model can effectively
reproduce the fine structure and width of the {\it central} part
of the FFP maxima becomes especially evident when looking at
$\delta=0.76$ where the side peaks are clearly separated and the
plateau-like structure of the central part is well
described, similar arguments apply to the wide maximum with the
slight dip for $\delta$=0.45. The power of our simple model is
also nicely illustrated when comparing the FFPs for $\delta=0.8$
and $0.85$ which are strikingly different despite the very similar
cavity geometry: For $\delta=0.8$, a distinct minimum in the
central plateau is evident in both experiment and theory
which is missing for $\delta=0.85$, again in both data sets.

{\it FFP and classical phase space} --- We now provide numerical
evidence that these distinct differences in the fine structure of
the FFP maxima originate in the structure of the underlying phase
space. We already discussed the Poincar\'{e} 
SOS of oval ({\it hard wall}) billiards in Fig.~\ref{newfig2}
and turn now to the properties of light leaving the {\it open}
system, cf.~Fig.~\ref{newfig5}. The Fresnel intensity transmitted
into far field direction $\phi_{\rm ff}$, 
originating from a certain near
field angle $\phi$ and an angle of incidence $\chi$ prior to transmission, is
represented in color scale histograms for the near and far field
intensity (lower and upper panels, respectively). To this end the
respective spaces, ($\sin\chi, \phi$)
and ($\phi_{\rm ff}, \phi$), 
were divided into cells in which the Fresnel intensity of
transmitted light rays was collected using the
ray model described above.

Superimposing the classical phase space from Fig.~\ref{newfig2} on
the light distribution in the near field (lower panels) reveals
that for all $\delta$ light rays leave indeed exclusively from the
high curvature regions. Moreover, the
larger $\delta$, the closer to the critical lines $|\sin\chi|
=1/n$ are the angles of incidence prior to transmission. That this
is a characteristic property of the underlying classical phase
space becomes even clearer when comparing the signatures in
Fig.~\ref{newfig5} with the fingerprint of light rays started at
the high curvature region with near-critical incidence in the
Poincar\'{e} SOS, see the lighter points in
Fig.~\ref{newfig2}.

For $\delta$=0.45, the structure in the emmiting regions is richer
than Fig.~\ref{newfig2} would suggest. This is related to the
intricate interplay between geometry, trajectory, and Fresnel's law
and illustrated in the
inset of Fig.~\ref{newfig5}. To this end 50 rays (marked by white
circles and numbered 0) leaving the cavity in the high intensity
region close to the critical line were followed for the next
reflections (marked by numbers, colors and white trajectory are
guides for the eye); those rays emit only little intensity at the first
bounces. Intensity is again emitted
refractively at the next subcritical reflections occuring 6 and 
7 bounces later (arrows in inset). The Poincar\'{e} SOS signature
of these bounces is marked by green squares (the other numbers
complete the Poincar\'{e} SOS schematically): 
the emergence of the line-like structure in the near field Fresnel
intensity (originating at bounce 7) becomes evident (symmetry
considerations complete the line pattern). Note that trajectories
started in the intensity gaps couple out refractively after few
bounces but not once the steady phase space distribution is reached.

The upper panels of Fig.~\ref{newfig5} show how the near field
patterns are translated into the far fields, the correspondence is
evident. Since the FFPs shown in Fig.~\ref{newfig4} are obtained
by summing the Fresnel intensities in the upper panels
horizontally (over all $\phi$), 
we conclude that the
observed far-field characteristics 
of (even) the lasing cavity is intimately
connected with, and can nicely be explained by, the {\it
classical} phase-space structure. Moreover, origin and evolution
of the fine structure of the FFP maxima in Fig.~\ref{newfig4} are
easily understood in the representation of Fig.~\ref{newfig5}:
Each FFP maxima in Fig.~\ref{newfig4} consists of two
contributions corresponding to light leaving from the two regions
of highest curvature. For $0.45
\leq \delta \leq 0.8$, the two contributions are separated in
$\phi_{\rm ff}$ 
causing a multipeak-structure of the FFPs, whereas for the
only slightly larger shape parameter $\delta=0.85$ they (almost)
merge, in excellent agreement with experiment.


{\it Conclusion} --- We fabricated oval-shaped resonant
microcavity lasers of various shapes, and lasing operation was
successfully established. We showed that the emitted 
{\it coherent} light
uniquely reflects the characteristics of the 
underlying {\it classical} phase space. We
employed a ray-based model 
that allows us to relate 
the observed peaks in the FFPs convincingly and unambiguously
to the phase-space structure. In particular we find that
the formation of a transversal barrier in the underlying classical
phase space around $\delta \lesssim 0.76$ can be detected by
coherent light.

We thank Susumu Shinohara and Satoshi Sunada for helpful
discussions. M.H. thanks the group of T.H. for warm hospitality
and support. The work at ATR was supported in part by the National
Institute of information and Communication Technology of Japan.



\begin{thebibliography}{99}
\bibitem{remark_FSR} This is commensurate with the free spectral range being about
1/100 of the observed lasing mode line width.
\bibitem{remark_pol} We assume transverse electric (TE) polarisation of light leaving
the microcavity.
\end{thebibliography}
\end{document}